\documentstyle[preprint,aps]{revtex}
\tightenlines
\begin{document}
\draft

\global \advance \count0 by -1

\thanks{This manuscript has been authored under contract number
DE-AC02-98CH10886 with the U.S.~Department of Energy.  Accordingly,
the U.S. Government retains a non-exclusive, royalty-free license to
publish or reproduce the published form of this contribution, or allow
others to do so, for U.S.~Government purposes.}

\title{Evaluating Grassmann Integrals}
\author {Michael Creutz}
\address{Physics Department, Brookhaven National Laboratory,
Upton, NY 11973, email: creutz@bnl.gov}

\maketitle

\begin{abstract}
I discuss a simple numerical algorithm for the direct evaluation of
multiple Grassmann integrals.  The approach is exact, suffers no
Fermion sign problems, and allows arbitrarily complicated
interactions.  Memory requirements grow exponentially with the
interaction range and the transverse size of the system.  Low
dimensional systems of order a thousand Grassmann variables can be
evaluated on a workstation.  The technique is illustrated with a
spinless fermion hopping along a one dimensional chain.
\end{abstract}


\input epsf

In path integral formulations of quantum field theory, fermions are
treated via integrals over anti-commuting Grassmann variables
\cite{grassmann}.  This gives an elegant framework for the formal 
establishment of Feynman perturbation theory.  For non-perturbative
approaches, such as Monte Carlo studies with a discrete lattice
regulator, these variables are more problematic.  Essentially all such
approaches formally integrate the fermionic fields in terms of
determinants depending only on the bosonic degrees of freedom.
Further manipulations give rise to the algorithms which dominate
current lattice gauge simulations.

Frequently, however, this approach has serious shortcomings.  In
particular, when a background fermion density is desired, as for
baryon rich regions of heavy ion scattering, these determinants are
not positive, making Monte Carlo evaluations tedious on any but the
smallest systems\cite{barbour}.  This problem also appears in studies of many
electron systems, particularly when doped away from half filling.

In this note I explore an alternative possibility of directly
evaluating the fermionic integrals, doing the necessary combinatorics
on a computer.  This is inevitably a rather tedious task, with the
expected effort growing exponentially with volume.  Nevertheless, in
the presence of the sign problem, all other known algorithms are also
exponential.  My main result is that this growth can be controlled to
a transverse section of the system.  I illustrate the technique with
low dimensional systems involving of order a thousand Grassmann
variables.

I begin with a set of $n$ anti-commuting Grassmann variables
$\{\psi_i\}$, satisfying
$[\psi_i,\psi_j]_+=\psi_i\psi_j+\psi_j\psi_i=0$. Integration is
uniquely determined up to an overall normalization by requiring
linearity and ``translation'' invariance {$\int d\psi
f(\psi+\psi^\prime) = \int d\psi f(\psi)$}.  For a single variable, I
normalize things so that
\begin{equation}
\int d\psi\ \psi = 1 \qquad \int d\psi\ 1 = 0.
\label{basic}
\end{equation}
In summary, to obtain a non-vanishing contribution, every integration
variable must appear exactly once in the expansion of the integrand.

Consider an arbitrary action $S(\psi)$, a function of these
anti-commuting variables, inserted into a path integral.  In
particular, I want to evaluate
\begin{equation}
Z=\int d\psi_n \ldots d\psi_1\ e^{S(\psi)}.
\label{pathintegral}
\end{equation}
Formally this requires expanding the exponent and keeping those terms
containing exactly one factor of each $\psi$.

I first convert the required expansion into operator manipulations in
a Fock space.  Introduce a fermionic creation-annihilation pair for
each fermionic field, {\it i.e.} $\psi_i \leftrightarrow
\{a_i^\dagger,a_i\}$.  These satisfy the usual relations
\begin{equation}
[a_i,a_j^\dagger]_+=\delta_{ij}.
\label{commutation}
\end{equation}
The space is built up by applying creation operators to the
vacuum, which satisfies $a_i{|0\rangle}=0$.  It is convenient
to introduce the completely occupied
``full'' state
\begin{equation}
{|F\rangle} \equiv a_n^\dagger...a_2^\dagger a_1^\dagger {|0\rangle}.
\label{full}
\end{equation}
With these definitions, I rewrite my basic path integral
as the matrix element
\begin{equation}
Z\ =\ {\langle 0|}\  e^{S(a)}\  {|F\rangle}.
\label{me1}
\end{equation}
Expanding {$e^S$}, a non-vanishing contribution requires one factor of
{$a_i$} for each Fermion.  This is the same rule as for Grassmann
integration.

I now manipulate this expression towards a sequential evaluation.
Select a single variable $\psi_i$ and define $S_i(a)$ as all terms
from the action involving a factor of $a_i$.  I define the complement
$\tilde S_i$ to be all other terms, so that $S=S_i+\tilde S_i$.  For
simplicity I assume that my action is bosonic so that {$S_i$} and
{$\tilde S_i$} commute.  Thus I have
\begin{equation}
Z={\langle 0 |} e^{\tilde S_i} e^{S_i}{|F\rangle}.
\label{me2}
\end{equation}
Observe that since $\tilde S_i$ contains no factors of $a_i$, the
occupation number for that variable, $n_i=a_i^\dagger a_i$, would
vanish if inserted between the two operators in Eq.\ (\ref{me2}).  I
thus can insert a projection operator $1-n_i$ between these factors
\begin{equation}
Z={\langle 0 |} e^{\tilde S_i}\ (1-n_i)\ e^{S_i}{|F\rangle}.
\label{me3}
\end{equation}
The next two steps are not essential to the algorithm, but simplify
the appearance of the final result.  First I use $a_i^2=0$ and the
fact that $S_i$ is linear in $a_i$.  Thus, the right hand factor
expands as only two terms
\begin{equation}
Z={\langle 0 |} e^{\tilde S_i}\ (1-n_i)\ (1+S_i){|F\rangle}.
\label{me4}
\end{equation}
Since $1-n_i$ projects out an empty state at
location $i$, I trivially have $a_i\ (1-n_i)=0$.  This means $S_i\
(1-n_i)=0$ and I can replace {$\tilde S_i$} on the left with the
full action
\begin{equation}
Z={\langle 0 |} e^{S}\ (1-n_i)\ (1+S_i){|F\rangle}.
\label{me5}
\end{equation}
Finally, I repeat this procedure successively for all variables,
giving the main result
\begin{equation}
Z=
\left\langle 0 \left\vert
 \prod_i \left( (1-n_i) (1+S_i)\right )
\right\vert F\right\rangle.
\label{central}
\end{equation}
where I have assumed that the action has any remaining constant pieces
removed.

This summarizes the basic algorithm.  Begin by setting up an
associative array for storing general states of the Fock space.
Standard hash table techniques allow rapid storage and retrieval.
More explicitly, for a given state
\begin{equation}
\vert \psi \rangle=\sum_s \chi_s |s\rangle.
\label{fock}
\end{equation}
store the numbers $\chi_s$ labeled by the respective states
$|s\rangle$.  At the outset this table is very short, only containing
one entry for the full state.  The algorithm proceeds with a loop over
the Grassmann variables.  For a given $\psi_i$, first apply $(1+S_i)$
to the stored state.  Then empty that location with the projector
$1-n_i$.  After all sites are integrated over, only the empty state
survives, with the desired integral as its coefficient.

\begin{figure}
\epsfxsize .5\hsize
\centerline {\epsfbox{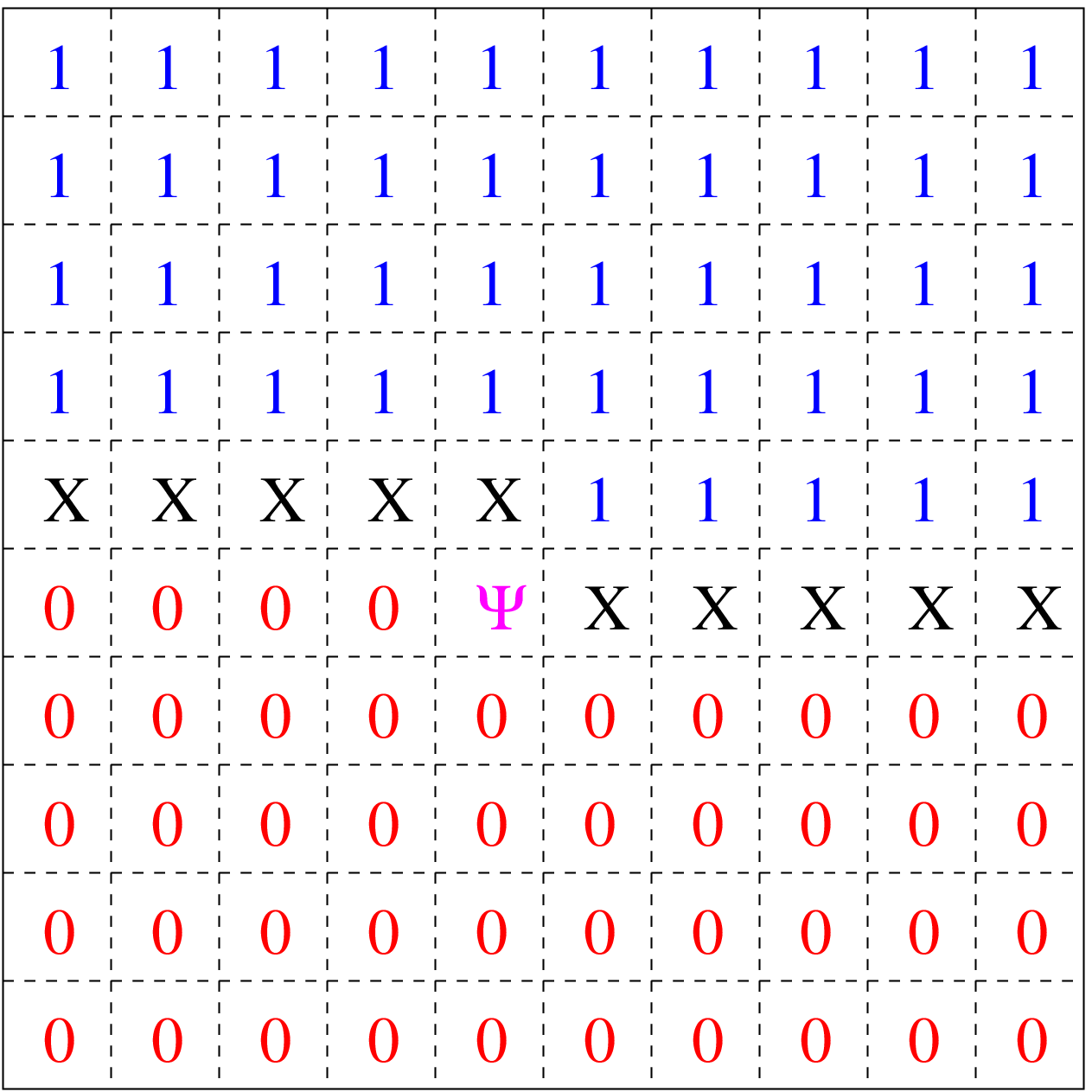}}
\smallskip
\caption {Integrating out sequentially, all finished sites are empty
and out of range sites are filled.  When integrating the site labeled
$\psi$, only those sites labeled ``X'' are undetermined.}
\label{latfig}
\end{figure}

The advantage of the scheme becomes apparent with a local interaction.
All sites that have previously been visited are empty, and thus
involve no information.  Any unvisited locations outside the range of
the interaction are still filled, and thus also involve no storage.
All relevant Fock states are nontrivial only for unvisited sites
within the interaction range of previously visited sites.  Sweeping
through the system in a direction referred to as ``longitudinal,'' we
only need keep track of a ``transverse'' slice of the model.  This is
illustrated in Fig.~(\ref{latfig}).  Thus, although the total number
of basis states in the Fock space is two to the number of Grassmann
variables, the storage requirements only grow as two to the transverse
volume of the system.

Note that the algorithm makes no assumptions about the precise form of
the interaction.  The approach is exact, with no sign problems.  The
complexity does grow severely with interaction range, probably
limiting practical applications to short range interactions in low
dimensions.  Note that the effort only grows linearly with the
longitudinal dimension, allowing very long systems in one direction.
This discussion has been in the context of ``real'' Grassmann
variables.  For ``complex'' variables simply treat {$\psi$} and
{$\psi^*$} independently.

In the transverse direction the boundary conditions are essentially
arbitrary, but longitudinal boundaries should not be periodic.  To
make them so requires maintaining information on both the top and
bottom layers of the growing integration region, squaring the
difficulty.  Note that the technique is similar to the finite lattice
method used for series expansions \cite{series}, and closely related
to attempts to directly enumerate fermionic world lines
\cite{worldline}.

\begin{figure}
\epsfxsize .7\hsize
\centerline {\epsfbox{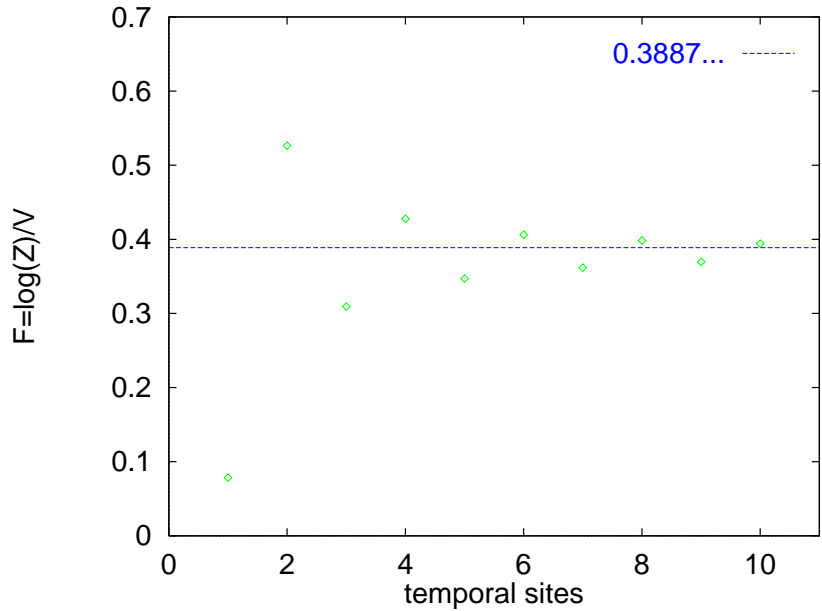}}
\smallskip
\caption {The free energy $F=\log(Z)/ N_i N_t$ 
on an $N_i=50$ site chain as a function of the number of time slices
$N_t$.  The infinite volume solution is shown by the horizontal line.}
\label{hopfig}
\end{figure}

For a simple demonstration, consider a spin-less fermion hopping along
a line of sites.  I introduce a complex Grassmann variable on each
site of a two dimensional lattice and study
\begin{equation}
Z=\int (d\psi d\psi^*) e^{S_t+S_h}
\end{equation}
with the temporal hopping term of form
\begin{equation}
S_t=\sum_{i,t} \psi^*_{i,t}(\psi_{i,t}-\psi_{i,t-1})
\end{equation}
and the spatial hopping 
\begin{equation}
S_h=k\sum_{i,t} \psi^*_{i,t}\psi_{i+1,t}+\psi^*_{i+1,t}\psi_{i,t}.
\end{equation}
I take $N_t$ sites in the time $t$ direction and $N_i$ spatial sites.
Here the one sided form of the temporal hopping insures the model has
an Hermitian transfer matrix in this direction\cite{transfer}.  I
treat the time direction as my ``transverse'' coordinate, growing the
lattice along the spatial chain.  After each row with $N_t=10$, the
number of Fock states stored in the hash table rises to 184,756.  In
Fig.~(\ref{hopfig}) I plot the resulting ``free energy'' $F=\log(Z)/
N_i N_t$ for $k=1$ on a 50 site chain as a function of the number of
time slices.  The 50 by 10 case has one thousand Grassmann variables.
This model is easily solved in the infinite length limit by Fourier
transform.  For infinite $N_t$ this gives
\begin{equation}
{\log(Z)\over N_i N_t}\rightarrow
\int_{-\pi\over 2}^{\pi\over 2} dq \log(1+2\cos(q))={0.388748\ldots}
\label{integral}
\end{equation}
Note how the results in the figure oscillate about this line,
indicating that the transfer matrix, while Hermitian, is not positive
definite.  The transfer matrix with this action becomes positive
definite only for $|k|<{\scriptstyle 1\over 2}.$

\begin{figure}
\epsfxsize .7\hsize
\centerline {\epsfbox{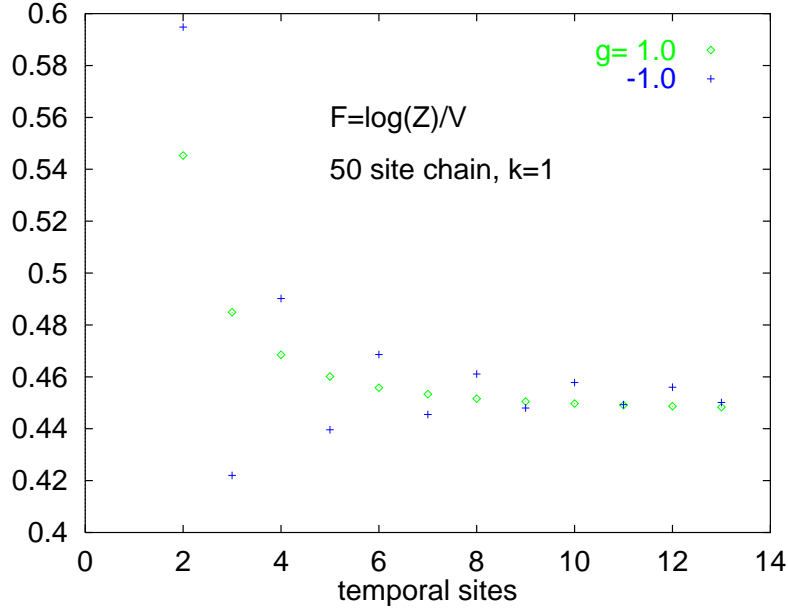}}
\smallskip
\caption {The free energy $F=\log(Z)/ N_i N_t$ with a four fermion
interaction as described in the text, plotted as a function of the
number of time slices $N_t$.  The chain has $N_i=50$ sites.  Points
are shown for $k=1$ and $g=\pm 1$.}
\label{lambdafig}
\end{figure}

Now I make the model somewhat less trivial with a four Fermion
interaction.  I take $S=S_t+ S_h+ S_I$ with
\begin{equation}
S_I=g\sum_{i,t} \psi^*_{i,t}\psi_{i,t}\psi^*_{i+1,t}\psi_{i+1,t}.
\end{equation}
In a transfer matrix formalism this represents an interaction
Hamiltonian of form {$H_I=\sum_i n_i n_{i+1}$}.  Bosonization relates
this Hamiltonian to the anisotropic quantum Heisenberg model, but this
equivalence is not being used here.  The above algorithm requires
essentially the same computer resources as the free case.  In
Fig.~(\ref{lambdafig}) I show the $N_t$ dependence for the free energy
with $k=1$ and $g=\pm 1$.  To reach $N_t=13$ for this figure, I
reduced memory requirements further by using temporal translation
invariance after each layer was integrated.

\begin{figure}
\epsfxsize .7\hsize
\centerline {\epsfbox{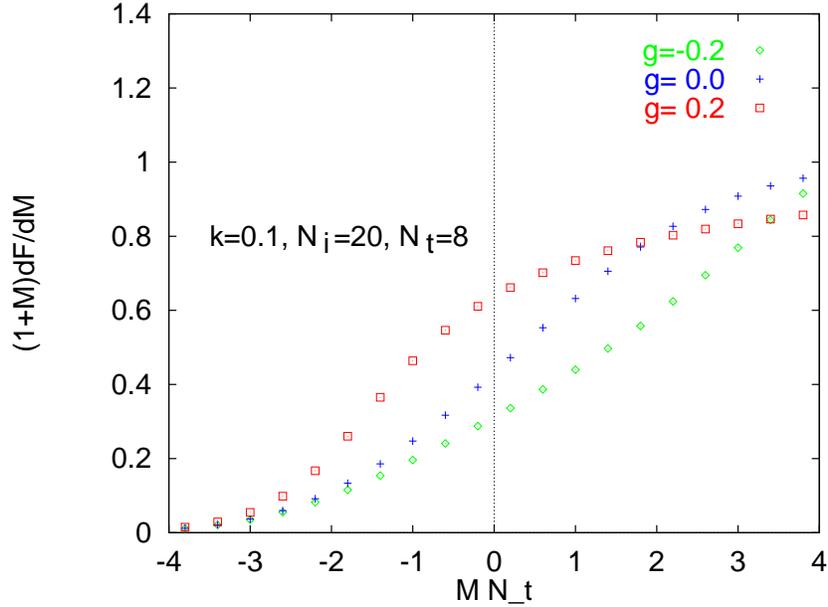}}
\smallskip
\caption {The occupancy of a 20 site chain as a function
of the chemical potential scaled by the number of time slices.  Note
how the filling occurs earlier or later depending on the sign of the
four fermion coupling.  Here the spatial hopping parameter is taken as
$K=0.1$}
\label{graph}
\end{figure}

With Monte Carlo studies of many fermion systems, the introduction of a
chemical potential term can be highly problematic due to
cancellations.  Here, however, it is just another local interaction of
negligible cost.  As an illustration, to the above model I add another
term and take $S=S_t+ S_h+ S_I+ S_M$ with
\begin{equation}
S_M=M\sum_{i,t} \psi^*_{i,t}\psi_{i,t}.
\end{equation}
I can use this term to regulate the ``filling,'' which can be
approximately monitored as $(1+M) {dF\over dM}$.  Here I include
the extra factor of $1+M$ to compensate partially for finite $N_t$
artifacts.  This quantity is shown as a function of $MN_t$ in
Fig.~(\ref{graph}), obtained on an $N_t=8$ by $N_i=20$ lattice with a
spatial hopping parameter of $k=0.1$.  For this figure I made a crude
extrapolation in chain length by defining $F={1\over
N_t}\log\left({Z(N_i)\over Z(N_i-1)}\right)$ .  Note how the four
fermion coupling enhances the filling.  The crossing of the curves at
large chemical potential is a consequence of strong coupling at finite
$N_t$.

An obvious system for future study is the Hubbard model\cite{hubbard}.
This requires 4 Grassmann variables per site corresponding to $\psi^*$
and $\psi$ for spins up and down.  Higher spatial dimensions strongly
increase the size of the transverse volume and will limit practical
system volumes, but this may be compensated for by the lack of sign
problems.

Another potential application relates to the use of Shockley
surface states to formulate chiral gauge theories in lattice gauge
theory.  These so called ``domain wall Fermions'' have unresolved
questions due to a natural pairing of surfaces.  One suggestion
\cite{chiral} proposes a four fermion coupling on one surface
to remove the spurious modes.  The effects of such a coupling might be
studied in a truncated model via the above techniques.

\end{document}